# How to study the evolution of parametrized games?

G. Karev

National Center for Biotechnology Information, National Institutes of Health, Bethesda, MD 20894, USA (retired); gkarev@gmail.com

## Introduction

Consider a family of games with corresponding payoff matrices $G^\beta$, parametrized by the parameter $\beta \in$ B. The informal question of interest is: "what game is the best?"

Surely, one should define what "best" means. Instead of attempting to suggest a (more or less arbitrary) definition of the "best" game, I allow natural selection to find it through "honest competition" between games. Specifically, I consider a model of a community composed of populations with different payoff matrices $G^\beta$ and study the natural selection between the populations. This process can be formally described by the evolution of the $\beta$ distribution. The outcome of the community evolution determines the "best" game.

Below I develop a mathematical "toolbox" (see Equations (3.11), (3.12)), that allows the study of evolution of parametrized games. The developed approach is applied to three examples of parametrized games from the literature: local replicator dynamics (C. Hilbe, 2011), pairwise competition (J. Morgan, K. Steglitz, 2003) and the game of alleles in diploid genomes (Bohl et.al, 2014)

## 1. Classical Replicator Equation and 2-Strategy Games

Suppose that individuals in a population can choose among $n$ strategies, and let $l_i(t)$ be the number of individuals ($i$-players) adopting the $i$-th strategy. Let an $i$-player interacting with a $j$-player obtain a payoff $g_{ij}$ and let $G = \left(g_{ij}\right), i, j = 1 \dots n$ be the payoff matrix.

Let us denote $N(t) = \sum_{i=1}^n l_i$ the total population size and $x_i = l_i/N$ the frequency of $i$-players. If the population is infinite and well-mixed, then the expected fitness of strategy $i$ (or the payoff obtained by the $i$-th strategist after interaction with the total population) is $\sum_{j=1}^n g_{ij} x_j = (Gx)_i$, so

$$\frac{dl_i}{dt} = l_i (Gx)_i. \qquad (1.1)$$

It follows from equation (1.1), that the frequencies of $i$-players solve the replicator equation

$$\frac{dx_i}{dt} = x_i[(Gx)_i - xGx].\qquad(1.2)$$

Here $xGx = \sum_{i,j=1}^{n} x_i g_{ij} x_j$ is the average payoff over the whole population (see Taylor and Jonker, 1978 for details).

In what follows, we consider 2-player games with 2 possible strategies, $s_0$ and $s_1$ (e.g., cooperators and defectors). Denote $x(t)$ the frequency of $s_1$ strategy. It is known (see, e.g., Hofbauer and Sigmund, 1998) that there exist 4 types of such games defined by the payoff matrices

$$G = \begin{bmatrix} a & b \\ c & d \end{bmatrix}.$$

Let

$$A = a - c, \ B = d - b.$$

Then equation (1.2) can be rewritten as

$$\frac{dx}{dt} = x(1-x)[x(A+B) - A].\qquad(1.3)$$

If $A < 0$ and $B > 0$, then the corresponding game is Prisoner's Dilemma (PD), where defectors dominate over cooperators; $x = 1$ is a stable equilibrium.

If $A > 0$ and $B < 0$, then the corresponding game is Harmony (H), where cooperators dominate over defectors; $x = 0$ is a stable equilibrium.

In both cases, there are no other stables states.

If both $A$ and $B$ are positive, then the game becomes Stag Hunt (SH) or coordination game. It is characterized by bi-stability, when both equilibria $x = 0$ and $x = 1$ are stable, and their areas of attraction are divided by the unstable equilibrium $x^* = \frac{A}{A+B}$.

Finally, if both $A$ and $B$ are negative, then the corresponding game is the Hawk–Dove (HD) game, also known as snowdrift or the game of chicken. In this game, there exists internal (polymorphic) stable equilibrium $x^* = A/(A+B)$, and equilibria $x=0$ and $x=1$ are unstable.

Below we present some examples of parametrized games based on the games of described types.

## 2.1. Local Replicator Dynamics

Classical replicator dynamics assume that each individual interacts with a representative sample of the infinite population. C. Hilbe (2011) considered *local replicator dynamics*: while the population itself is infinite, interactions and reproduction occur in random groups of size $N \geq 2$. Hilbe proved that if groups are formed according to a multinomial distribution, then the local replicator dynamics is given by the *modified replicator equation*:

$$\frac{dx_i}{dt} = x_i[(\widetilde{G}\,x)_i - x(\widetilde{G}\,x)] \tag{2.1}$$

with $\widetilde{G} = G - \frac{G+G^T}{N}$.

Let us denote $\beta = \frac{1}{N}$, $0 < \beta \leq \frac{1}{2}$, and represent the matrix $\widetilde{G}$ in the form

$$\widetilde{G} = G - \frac{G+G^T}{N} = G - \beta(G + G^T),$$

Denote

$$G^\beta = G(1-\beta) - \beta G^T = \begin{bmatrix} a(1-2\beta) & b(1-\beta) - \beta c \\ c(1-\beta) - \beta b & d(1-2\beta) \end{bmatrix}. \tag{2.2}$$

Then, according to Hilbe (2011), local replicator dynamics is described by the replicator equation

$$\frac{dx_i}{dt} = x_i[(G^\beta x)_i - xG^\beta x] \tag{2.3}$$

with the parametrized payoff matrix given by (2.2).

### 2.2. Pairwise Competition

From the Summary of the paper by J. Morgan and K. Steglitz (2003):

"Spite in Hamilton's sense is defined as the willingness to harm oneself in order to harm another more. The standard replicator dynamic predicts that evolutionarily stable strategies are payoff-maximizing equilibria of the underlying game, and hence rules out the evolution of spiteful behavior. We propose a modified replicator dynamic, where selection is based on local outcomes, rather than on the population 'state', as in standard models."

Let $G = (g_{ij}), i, j = 1 \dots n$ be the payoff matrix of the underlying game. Then the spiteful replicator dynamic, according to Morgan & Steglitz, is described by the equation

$$\frac{dx_i}{dt} = x_i \sum_j (g_{ij} - g_{ji})x_j = x_i[((G - G^T)x)_i]. \tag{2.4}$$

To characterize intermediate situations between the standard and the spiteful replicator dynamic, Morgan & Steglitz introduced the payoff matrix

$$G^\beta = G - \beta G^T, \qquad 0 \le \beta \le 1$$

where the parameter $\beta$ characterizes the "degree of spite". In our notation, it is

$$G^\beta = \begin{bmatrix} a(1-\beta) & b - \beta c \\ c - \beta b & d(1-\beta) \end{bmatrix}. \tag{2.5}$$

Now the intermediate spiteful dynamics is described by the replicator equation:

$$\frac{dx_i}{dt} = x_i(\sum_j (g_{ij} - \beta g_{ji})x_j - (1-\beta)\sum_{i,j} g_{ij}x_i x_j) = x_i[(G^\beta x)_i - xG^\beta x] \ . \tag{2.6}$$

The question we are interested in is the following: what "degree of spite" will be selected depending on the underlying game with the payoff matrix $G$?

### 2.3. Alleles in Diploid Genomes Playing Two-Player Games

From Bohl et.al, "Evolutionary game theory: molecules as players", 2014:

> "Traulsen and Reed revisit the formerly rarely considered interpretation of interactions between alleles in a diploid genome as a two-player game. For the example of "meiotic drive", where a gene is said to "drive" when being transmitted at a higher probability than fair 50% "

Traulsen and Reed set up the following payoff matrix (in our notations):

$$G^\beta = \begin{matrix} & \begin{matrix} S & \ \ \ D \end{matrix} \\ \begin{matrix} S \\ D \end{matrix} & \begin{bmatrix} a & b(1-\beta) \\ b\beta & d \end{bmatrix} \end{matrix} \tag{2.7}$$

where $\beta$ is the probability that the driving allele $D$ is transferred from a heterozygous parent to the offspring instead of the susceptible allele $S$; $a$, $b$ and $d$ are the relative finesses of the different genotypes over an entire life cycle.

Four different situations arise corresponding to four distinct games: (I) a Prisoner's Dilemma when the drive allele can invade and reach fixation, (II) a snowdrift game when the allele can invade but not reach fixation, (III) a harmony game when the allele can neither invade nor reach fixation and (IV) a coordination game when the allele cannot invade but can reach fixation when starting at a sufficiently high initial frequency."

## 3. Mathematical "Toolbox": How to Study Evolution of the Parameter Distribution in Parametrized Games

Let the parametrized payoff matrix $G^\beta$ be given. Let $l(t; \alpha, \beta)$ be the density of a clone in $\beta$-population, which adopts the $s_1$ strategy with probability $\alpha$ at time $t$. Let us denote $x(t)$ as the frequency of $s_1$-players in the total community at time $t$.

The players from the $\beta$-population that keep $s_0$- or $s_1$-strategies get the payoffs in time $t$ correspondingly:

$$f_0(\beta, x(t)) = G_{00}^{(\beta)}(1 - x(t)) + G_{01}^{(\beta)}x(t), \qquad (3.1)$$

$$f_1(\beta, x(t)) = G_{10}^{(\beta)}(1 - x(t)) + G_{11}^{(\beta)}x(t).$$

Then the dynamics of the clones is described by the equation:

$$\frac{dl(t; \alpha, \beta)}{dt} = l(t; \alpha, \beta)F(t; \alpha, \beta) \qquad (3.2)$$

where

$$F(t; \alpha, \beta) = \alpha f_1(\beta, x(t)) + (1 - \alpha)f_0(\beta, x(t)).$$

In more detail,

$$F(t; \alpha, \beta) = G_{00}^{(\beta)} + \alpha(G_{10}^{(\beta)} - G_{00}^{(\beta)}) + x(t)(G_{01}^{(\beta)} - G_{00}^{(\beta)}) + \alpha x(t)(G_{00}^{(\beta)} - G_{01}^{(\beta)} - G_{10}^{(\beta)} + G_{11}^{(\beta)}). \qquad (3.3)$$

Equation (3.2), (3.3) can be solved by the HKV-method (see Karev, 2010, 2012; Kareva and Karev, 2020) as follows.

Introduce formally the auxiliary keystone variable $q(t)$ using the following "escort equation":

$$\frac{dq(t)}{dt} = x(t), q(0) = 0. \qquad (3.4)$$

Then we can write the solution to equation (3.2) in the form

$$l(t; \alpha, \beta) = l(0; \alpha, \beta)\exp(W_0 + \alpha W), \qquad (3.5)$$

where

$$W_0(t, \beta) = (-G_{00}^{(\beta)} + G_{01}^{(\beta)})q(t) + G_{00}^{(\beta)}t,$$

$$W(t, \beta) = (G_{00}^{(\beta)} - G_{01}^{(\beta)} - G_{10}^{(\beta)} + G_{11}^{(\beta)})q(t) + (G_{10}^{(\beta)} - G_{00}^{(\beta)})t.$$

The total community size is given by

$$N(t) = \iint_{A,B} l(t; \alpha, \beta)\, d\alpha d\beta = N(0)\iint_{A,B}\exp(W_0(t, \beta) + \alpha W(t, \beta))\, P(0, \alpha, \beta)d\alpha d\beta,$$

where $\alpha \in A$, $\beta \in B$, and initial distribution $P(0, \alpha, \beta)$ is supposed to be given.

The current pdf

$$P(t, \alpha, \beta) = \frac{l(t; \alpha, \beta)}{N(t)} = P(0, \alpha, \beta) \frac{\exp(W_0(t,\beta) + \alpha W(t,\beta))}{\iint_{A,B} \exp(W_0(t,\beta) + \alpha W(t,\beta)) P(0,\alpha,\beta) \, d\alpha d\beta} \ . \qquad (3.6)$$

Next, the number of players keeping the $s_1$-strategy is given by:

$$D(t) = \iint_{A,B} \alpha \, l(t; \alpha, \beta) \, d\alpha d\beta =$$

$$N(0) \iint_{A,B} \alpha \exp(W_0(t,\beta) + \alpha W(t,\beta)) P(0; \alpha, \beta) \, d\alpha d\beta \ .$$

The frequency of $s_1$-strategy players is:

$$x(t) = \frac{D(t)}{N(t)} = \frac{\iint_{A,B} \alpha \exp(W_0(t,\beta) + \alpha W(t,\beta)) P(0;\alpha,\beta) \, d\alpha d\beta}{\iint_{A,B} \exp(W_0(t,\beta) + \alpha W(t,\beta)) P(0,\alpha,\beta) \, d\alpha d\beta} = E^t[\,\alpha\,]. \qquad (3.7)$$

Using equation (3.7), we can obtain an explicit equation for the auxiliary variable $q$, defined by equation (3.4). Having the solution $q(t)$, we can compute the pdf $P(t, \alpha, \beta)$ and all the statistical characteristics of interest in the model.

Natural selection of strategies, i.e., evolution of the distribution of parameter $\alpha$, was studied in (Karev, 2018). Here, we are interested mostly in the dynamics of parameter $\beta$, so in what follows, we assume for simplicity that there are no mixed strategies, $\alpha$ is independent of parameter $\beta$ and may take only two values, 0 and 1, with corresponding probabilities $p_0$ and $p_1$, at the initial time.

Let us denote

$$W_1(t, \beta) = W_0(t, \beta) + W(t, \beta) = \left(-G_{10}{}^{(\beta)} + G_{11}{}^{(\beta)}\right) q(t) + G_{10}{}^{(\beta)} t \ . \qquad (3.8)$$

Recall that

$$W_0(t, \beta) = \left(-G_{00}{}^{(\beta)} + G_{01}{}^{(\beta)}\right) q(t) + G_{00}{}^{(\beta)} t. \qquad (3.9)$$

Let us denote

$$V_0(t) = \int_B \exp(W_0(t,\beta)) P(0,\beta) d\beta \ , \qquad (3.10)$$

$$V_1(t) = \int_B \exp(W_1(t,\beta)) P(0;\beta) d\beta. \qquad$$

Then equations (3.4), (3.6) read:

$$\frac{dq}{dt} = \frac{p_1 V_1(t)}{p_0 V_0(t) + p_1 V_1(t)}, \qquad (3.11)$$

$$P(t, \beta) = P(0, \beta) \left( \frac{p_0 \exp(W_0(t,\beta)) + p_1 \exp(W_1(t,\beta))}{p_0 V_0(t) + p_1 V_1(t)} \right). \qquad (3.12)$$

Equations (3.11), (3.12), together with equations (3.8) - (3.10), compose the "toolbox" for studying the evolution of the parameter $\beta$ distribution (initial version of this toolbox was given in Kareva, Karev, 2019).

Now let us apply the developed "toolbox" to several specific problems.

## 4. Local Replicator Dynamics

Let $G$ be a payoff matrix for a game; then according to Hilbe (2011), see Equations (2.2), (2.3) above, corresponding parametrized matrix for local replicator dynamics is:

$$G^\beta = G(1-\beta) - \beta G^T, 0 < \beta \le 1/2.$$

If $G = \begin{bmatrix} a & b \\ c & d \end{bmatrix}$, then

$$G^\beta = \begin{bmatrix} a(1-2\beta) & b(1-\beta) - c\beta \\ c(1-\beta) - b\beta & d(1-2\beta) \end{bmatrix}. \tag{4.1}$$

Next, using equations (3.8), (3.9) and matrix (4.1), we get:

$$W_0(t,\beta) = (-a+b)q(t) + at + \beta(2a - b - c)q(t) - 2\beta at, \tag{4.2a}$$

$$W_1(t,\beta) = (-c+d)q(t) + ct + \beta(b + c - 2d)\,q(t) - \beta(b+c)t. \tag{4.2b}$$

Now we can solve equation (3.11) for the keystone variable $q(t)$ with $W_0, W_1$ given by equations (4.2). Having the solution $q(t)$, we can compute the distribution of the parameter $\beta$ at any time. Below some examples are given. In all examples in this section, the initial distribution $\Pr(0,\beta)$ of parameter $\beta$ is taken as exponential, truncated on the interval $(0,1/2)$:

$$\Pr(0,\beta) = \frac{1}{1-e^{-1/2}} e^{-\beta}, 0 < \beta \le 1/2.$$

Initial frequencies of strategies $s_0$ and $s_1$ are taken as $p_0 = p_1 = 0.5$.

Example 1.

Let $G = \begin{bmatrix} 2 & 0 \\ 3 & 1 \end{bmatrix}$. Then

$$G^\beta = \begin{bmatrix} a(1-2\beta) & b(1-\beta) - c\beta \\ c(1-\beta) - b\beta & d(1-2\beta) \end{bmatrix} = \begin{bmatrix} 2(1-2\beta) & -3\beta \\ 3(1-\beta) & (1-2\beta) \end{bmatrix}. \tag{4.3}$$

Evolution of parameter $\beta$ is shown in Fig.1. One can see that the distribution tends to be concentrated in point $\beta = 0$, indicating that in this case, natural selection favors the largest possible group size.

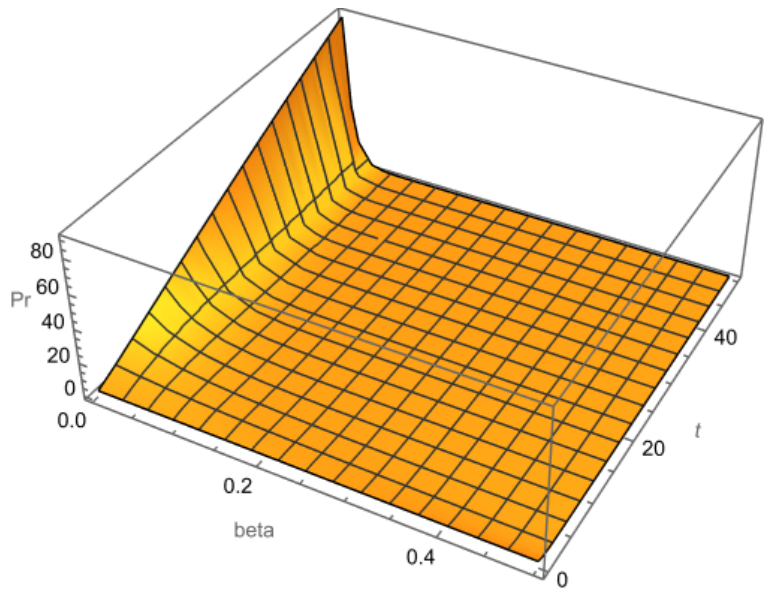

**Fig. 1**. Dynamics of the distribution of parameter $\beta$ for parametrized game (4.3). Natural selection favors the largest group sizes.

Example 2.

Let $G = \begin{bmatrix} -1 & -5 \\ 3 & -4 \end{bmatrix}$. Then

$$G^{\beta} = \begin{bmatrix} a(1-2\beta) & b(1-\beta) - c\beta \\ c(1-\beta) - b\beta & d(1-2\beta) \end{bmatrix} = \begin{bmatrix} -1+2\beta & -5+2\beta \\ 3+2\beta & -4+8\beta \end{bmatrix}. \tag{4.4}$$

Evolution of the distribution of parameter $\beta$ is shown in Fig.2. The distribution tends to be concentrated in the point $\beta = 1$, indicating that in this case, natural selection favors the smallest group size $N = 1$.

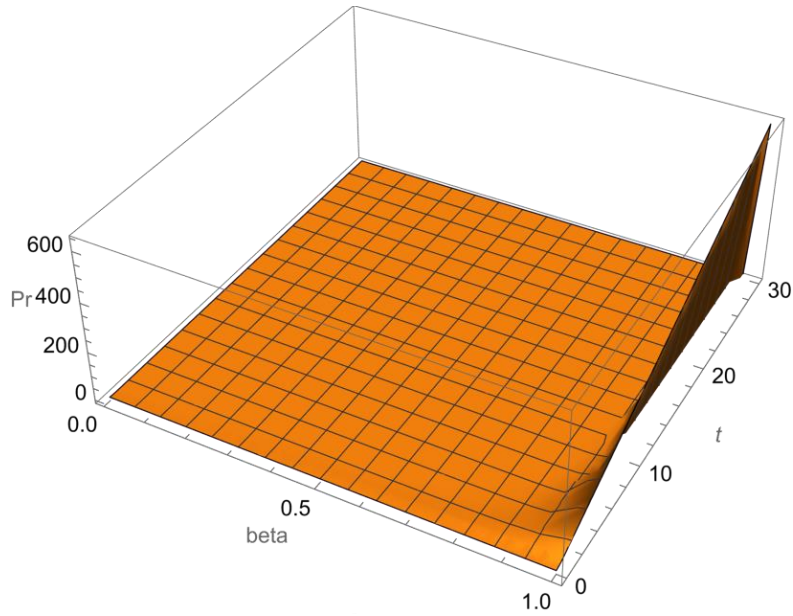

**Fig.2**. Dynamics of the distribution of parameter $\beta$ for parametrized game (4.4). Natural selection favors the smallest group sizes.

Example 3.

Let $G = \begin{bmatrix} -2 & 2 \\ 0 & 2.1 \end{bmatrix}$. Then

$$G^{\beta} = \begin{bmatrix} a(1-2\beta) & b(1-\beta)-c\beta \\ c(1-\beta)-b\beta & d(1-2\beta) \end{bmatrix} = \begin{bmatrix} -2(1-2\beta) & 2(1-\beta) \\ -2\beta & 2.1(1-2\beta) \end{bmatrix}. \tag{4.5}$$

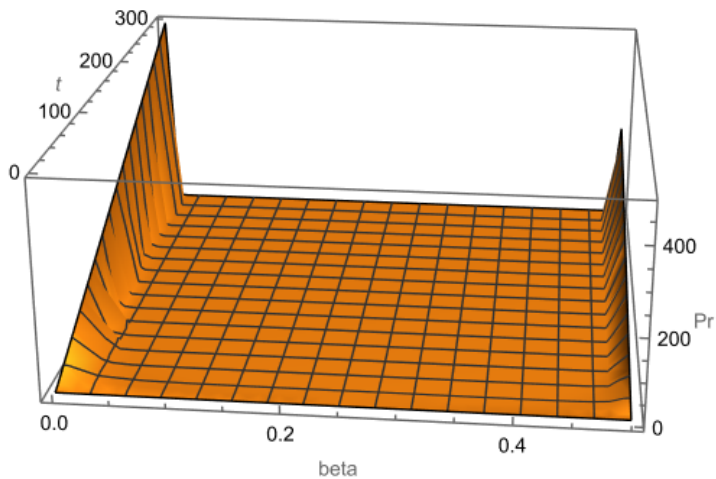

Fig.3. Dynamics of the distribution of parameter $\beta$ for parametrized game (4.5). Natural selection favors both the smallest least and the largest group sizes.

Example 4.

As the last example, consider the game with the payoff matrix

$G = \begin{bmatrix} 9 & -1 \\ 10 & 0 \end{bmatrix}$, then

$$G^\beta = \begin{bmatrix} a(1-2\beta) & b(1-\beta)-c\beta \\ c(1-\beta)-b\beta & d(1-2\beta) \end{bmatrix} = \begin{bmatrix} 9(1-2\beta) & -(1-\beta)-10\beta \\ 10(1-\beta)+\beta & 0 \end{bmatrix}. \qquad (4.6)$$

Evolution of the parameter $\beta$ distribution is shown on the Fig. 4.

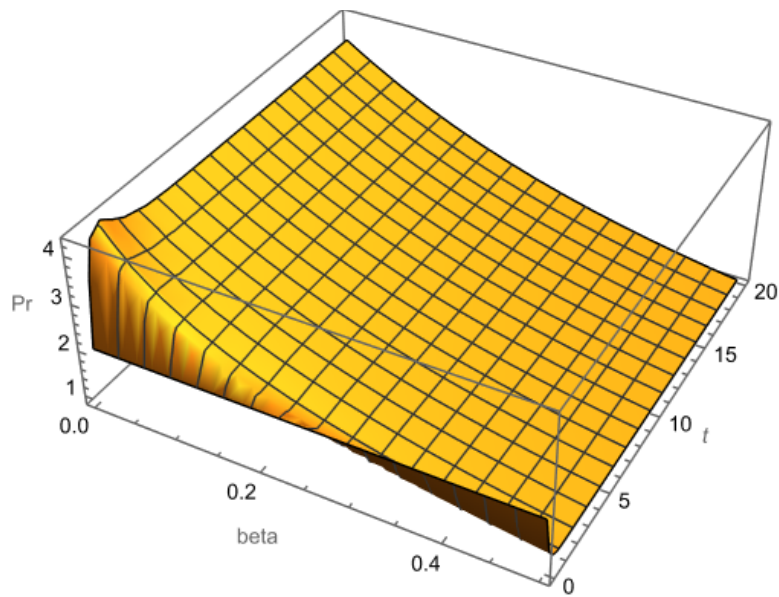

Fig.4. Dynamics of the distribution of parameter $\beta$ for parametrized game (4.6).

One can see that the distribution of parameter $\beta$ stays non-degenerated for all time, and natural selection preserves all $\beta$ -populations; it means that in this case, there is no "best" group size (although larger group sizes are "better").

These four examples show the main possibilities of natural selection between parametrized games; we will meet them in other examples below.

Notice that in all examples in this section the payoff matrices $G^\beta$ correspond to Prisoner's Dilemma game for all $\beta$; nevertheless, the results of natural selection are very different.

## 5. Pairwise Competition

Let $G = \begin{bmatrix} a & b \\ c & d \end{bmatrix}$ be the payoff matrix for the underlying game. If parameter $\beta$ characterizes the "degree of spite" (see s.2.2), then, according to Morgan & Steglitz (2003), the payoff matrix for pairwise competition replicator dynamics is $G^{\beta} = G - \beta G^T$, $0 \leq \beta \leq 1$, or

$$G^{\beta} = \begin{bmatrix} a(1-\beta) & b - \beta c \\ c - \beta b & d(1-\beta) \end{bmatrix}. \tag{5.1}$$

Then

$$W_0(t,\beta) = \left(-G_{00}{}^{(\beta)} + G_{01}{}^{(\beta)}\right) q(t) + G_{00}{}^{(\beta)} t = (b - \beta c - a(1-\beta))q(t) + a(1-\beta)t, \tag{5.2}$$

$$W_1(t,\beta) = \left(-G_{10}{}^{(\beta)} + G_{11}{}^{(\beta)}\right)q(t) + G_{10}{}^{(\beta)} t = \left(-c + \beta b + d(1-\beta)\right)q(t) + (c - \beta b)t.$$

Let us consider some examples of the "degree of spite" dynamics. In all examples in this and the next sections the initial distribution $\Pr(0,\beta)$ of parameter $\beta$ is taken as exponential, truncated on the interval $(0,1)$:

$$\Pr(0,\beta) = \frac{1}{1-e^{-1}} e^{-\beta}, 0 < \beta \leq 1.$$

Initial frequencies of strategies $s_0$ and $s_1$ are taken as $p_0 = p_1 = 0.5$.

Example 5. Let $G = \begin{bmatrix} 2 & 0 \\ 3 & 1 \end{bmatrix}$ be the payoff matrix of initial game; then the corresponding parametrized payoff matrix

$$G^{\beta} = G - \beta G^T = \begin{bmatrix} 2(1-\beta) & -3\beta \\ 3 & 1-\beta \end{bmatrix}. \tag{5.2}$$

Evolution of the distribution of parameter $\beta$ for parametrized game (5.2) is shown in Fig.5.

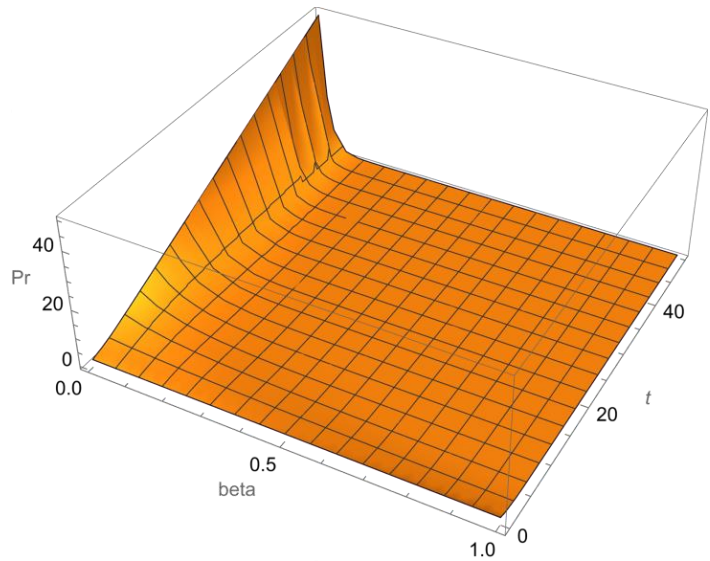

Fig. 5. Dynamics of the distribution of parameter $\beta$ for parametrized game (5.2).

One can see that in this case, natural selection favors the population with $\beta = 0$. In such a population, the individuals do not wish to harm themselves in order to harm others more.

Example 6. An opposite situation is illustrated by the following example.

Let $G = \begin{bmatrix} 3 & -3 \\ 4 & -2 \end{bmatrix}$. Then

$$G^\beta = G - \beta G^T = \begin{bmatrix} 3(1-\beta) & -3+2\beta \\ 4+3\beta & 2(1-\beta) \end{bmatrix}. \tag{5.3}$$

Evolution of the distribution of parameter $\beta$ for parametrized game (5.3) is shown in Fig.6.

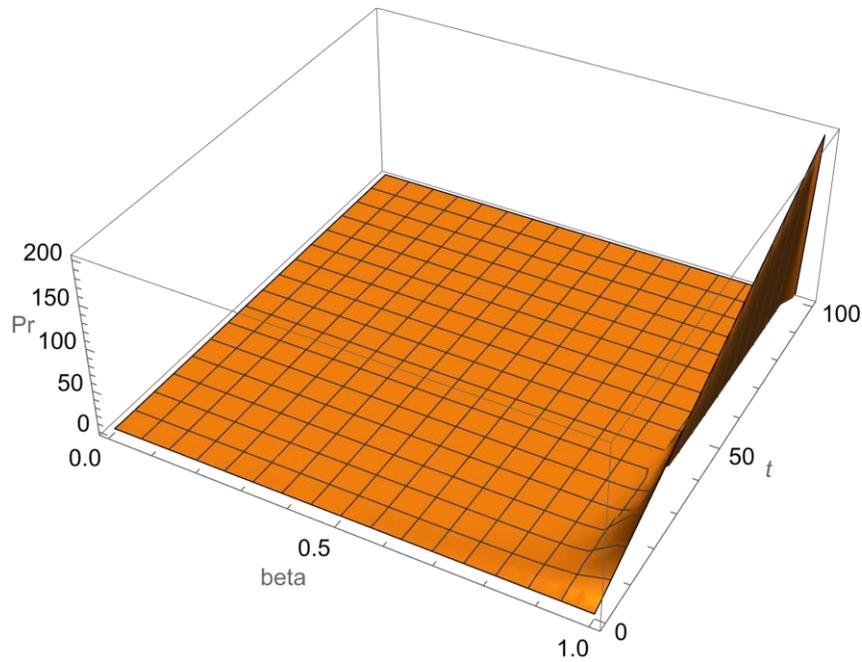

Fig.6. Dynamics of the distribution of parameter $\beta$ for parametrized game (5.3).

In the case of the parametrized game (5.3), natural selection favors the population with $\beta = 1$. In such a population, individuals are willing to harm themselves maximally to harm others even more.

Example 7. As the last example, consider the game with the payoff matrix $G = \begin{bmatrix} 1 & -5 \\ 3 & 1 \end{bmatrix}$. Then

$$G^{\beta} = G - \beta G^T = \begin{bmatrix} 1-\beta & -5-3\beta \\ 3+5\beta & 1-\beta \end{bmatrix}. \tag{5.4}$$

Evolution of the distribution of parameter $\beta$ for the parametrized game (5.4) is shown in Fig.6.

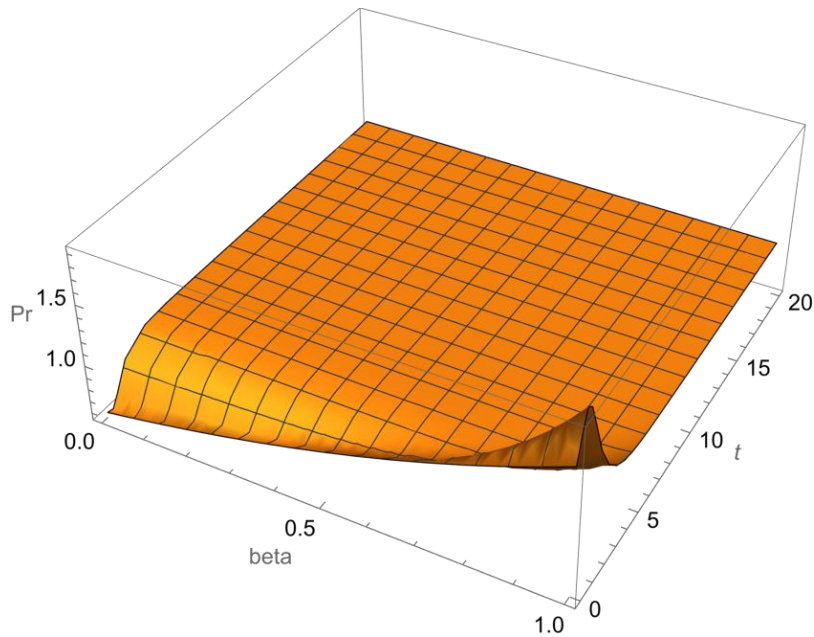

**Fig.6**. Dynamics of the distribution of parameter $\beta$ for the parametrized game (5.4).

In the case of parametrized game (5.4), natural selection does not favor any population with any specific value of the parameter $\beta$.

Notice that in all examples in this section the payoff matrices $G^\beta$ correspond to Prisoner's Dilemma game for all $\beta$, but the results of natural selection are different.

## 6. Games of alleles in diploid genomes

Traulsen and Reed (2011) set up the following payoff matrix (see s.2.3):

$$G^\beta = \begin{matrix} & S & D \\ S \\ D \end{matrix} \begin{bmatrix} a & b(1-\beta) \\ b\beta & d \end{bmatrix},$$

where $\beta$ is the probability that the driving allele $D$ is transferred from a heterozygous parent to the offspring instead of the susceptible allele $S$; $a$, $b$ and $d$ are the relative finesses of different genotypes over an entire life cycle. Notice that $G^1 = (G^0)^T$.

Examples 8,

Let $G = \begin{bmatrix} 1 & 2 \\ 2 & 1 \end{bmatrix}$, then

$$G^\beta = \begin{bmatrix} 1 & 2(1-\beta) \\ 2\beta & 1 \end{bmatrix}. \tag{6.1}$$

Evolution of the probability $\beta$ in this case is shown in Fig.7.

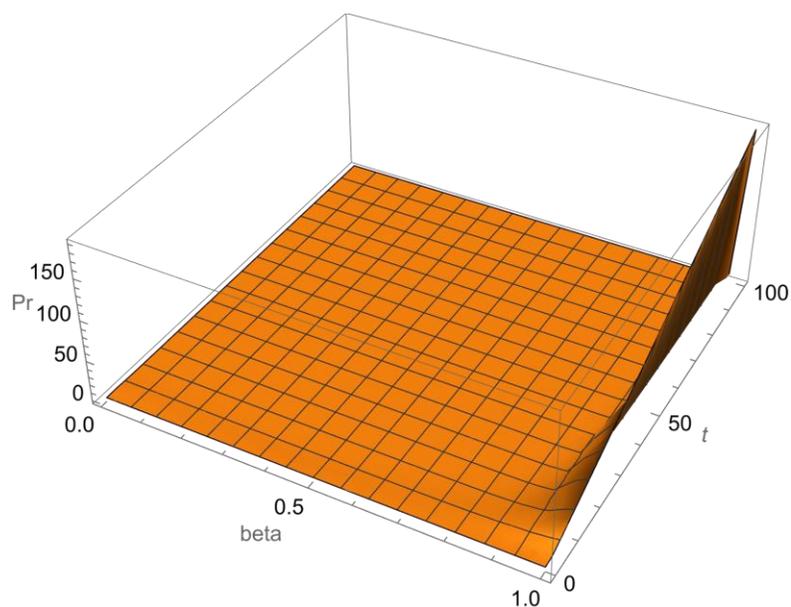

**Fig.7**. Evolution of the distribution of parameter $\beta$ for the parametrized game (6.1). Natural selection favors the population that has probability $\beta = 1$.

Example 9. Now let $G = \begin{bmatrix} -4 & -3 \\ 2 & -2 \end{bmatrix}$. Then $G^\beta = \begin{bmatrix} -4 & -3(1-\beta) \\ 2\beta & -2 \end{bmatrix}$. $\qquad$ (6.2)

Evolution of the probability $\beta$ in this case is shown in Fig. 8.

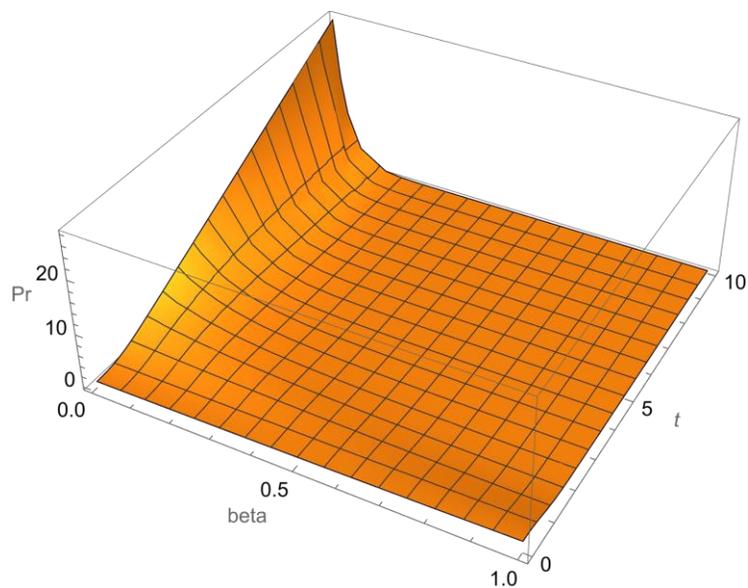

**Fig.8**. Evolution of the distribution of parameter $\beta$ for the parametrized game (6.2). Natural selection favors the population that has probability $\beta = 0$.

Example 10. As the last example, let $G = \begin{bmatrix} 1 & -5 \\ 3 & 1 \end{bmatrix}$, then

$$G^\beta = \begin{bmatrix} 1 & -5(1-\beta) \\ 3\beta & 1 \end{bmatrix}. \qquad (6.3)$$

Evolution of the probability $\beta$ in this case is shown in Fig.9. One can see that the distribution of $\beta$ stays non-degenerated at all times, and natural selection preserves all $\beta$ -populations.

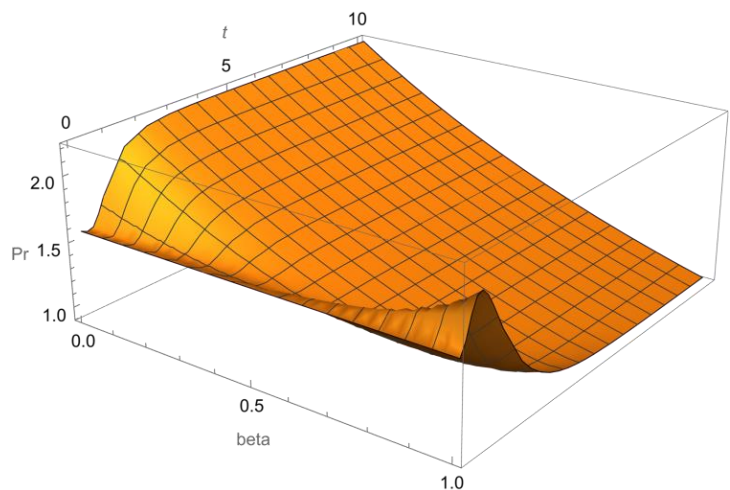

**Fig.9**. Evolution of the distribution of parameter $\beta$ for the parametrized game (6.3); natural selection preserves all $\beta$ - populations.

## Conclusion

In this paper I study the natural selection between parametrized games. Natural selection is described by the evolution of the distribution of the game parameter. The mathematical "toolbox" for studying the evolution of parameter distribution was developed and applied to three interesting parametrized games known from the literature, namely, local replicator dynamics (C. Hilbe, 2011), pairwise competition (J. Morgan, K. Steglitz, 2003) and the game of alleles in diploid genomes (A.Traulsen, F. Reed, 2011; Bohl et.al, 2014). Evolution of other parametrized games can be explored similarly.

**References**

K. Bohl, S. Hummert, S. Werner, D. Basanta, A. Deutsch, S. Schuster, G. Theißen and A. Schroeter (2014). Evolutionary game theory: molecules as players. Mol. BioSyst, 10, 3066.

C. Hilbe (2011). Local Replicator Dynamics: A Simple Link between Deterministic and Stochastic Models of Evolutionary Game Theory. Bull Math Biol. 73: 2068-2087

Hofbauer, J., & Sigmund, K. (1998). *Evolutionary games and population dynamics*. Cambridge university press.

Karev, G.P. (2010). On mathematical theory of selection: continuous time population dynamics. *J. Math. Biol. 60*, 107-129.

Karev, G.P. (2012).The HKV method of solving of replicator equations and models of biological populations and communities. arXiv:1211.6596v.

Karev, G. (2018). Evolutionary games: natural selection of strategies. arXiv:1802.07190.

Kareva, I., Karev, G. (2019). Natural selection between two games with applications to game theoretical models of cancer. *Bulletin of mathematical biology*, *81*, 2117-2132.

Kareva I.,  Karev G. (2020). *Modeling Evolution of Heterogeneous Populations. Theory and Applications*. Elsevier.

J. Morgan, K. Steiglitz (2003). Pairwise Competition and the Replicator Equation. *Bulletin of Mathematical Biology*  65, 1163–1172

P. D. Taylor, L. Jonker. (1978). Evolutionarily stable strategies and game dynamics. *Math. Biosci.* 40: 145–56.

A. Traulsen, F. Reed (2011). From genes to games: cooperation and cyclic dominance in meiotic drive. J. of Theor. Biology, 299, 120—5.